\documentclass[conference]{IEEEtran}
\IEEEoverridecommandlockouts
\usepackage{cite}
\usepackage{amsmath,amssymb,amsfonts}
\usepackage{algorithmic}
\usepackage{graphicx}
\usepackage{textcomp}
\usepackage{xcolor}
 \usepackage{float}
\usepackage{parskip}
\usepackage{caption}

\def\BibTeX{{\rm B\kern-.05em{\sc i\kern-.025em b}\kern-.08em
    T\kern-.1667em\lower.7ex\hbox{E}\kern-.125emX}}

\usepackage{url}

\begin{document}

\title{TX-Digital Twin: Visualizing Supercomputer GPU Performance Data Stream
\thanks{Research was sponsored by the Department of the Air Force Artificial
Intelligence Accelerator and was accomplished under Cooperative Agreement
Number FA8750-19-2-1000. The views and conclusions contained in this
document are those of the authors and should not be interpreted as representing
the official policies, either expressed or implied, of the Department of the
Air Force or the U.S. Government. The U.S. Government is authorized to
reproduce and distribute reprints for Government purposes notwithstanding
any copyright notation herein.}
}

%\author{\IEEEauthorblockN{Elena Baskakova, William Bergeron, Matthew Hubbell, Hayden Jananthan, Jeremy Kepner}
%\\ \vspace{-0.5cm}

\author{\IEEEauthorblockN{Elena Baskakova}
 \IEEEauthorblockA{\textit{MIT} \\
 Cambridge, USA \\
 ebask@mit.edu}
 \and
 \IEEEauthorblockN{William Bergeron}
 \IEEEauthorblockA{\textit{MIT} \\
 Cambridge, USA \\
 bbergeron@ll.mit.edu}
 \and
 \IEEEauthorblockN{Matthew Hubbell}
 \IEEEauthorblockA{\textit{MIT} \\
 Cambridge, USA \\
 mhubbell@ll.mit.edu}
 \and
 \IEEEauthorblockN{Hayden Jananthan}
 \IEEEauthorblockA{\textit{MIT} \\
 Cambridge, USA \\
 hayden.jananthan@ll.mit.edu}
 \and
 \IEEEauthorblockN{Jeremy Kepner}
 \IEEEauthorblockA{\textit{MIT} \\
 Cambridge, USA \\
 kepner@ll.mit.edu}
}

\IEEEoverridecommandlockouts
\IEEEpubid{\makebox[\columnwidth]{979-8-3315-5937-3/25/\$31.00 \copyright2025 IEEE \hfill} \hspace{\columnsep}\makebox[\columnwidth]{ }}
\maketitle
\IEEEpubidadjcol
\begin{abstract}
Supercomputers are complex, dynamic systems that serve thousands of users and are built with thousands of compute nodes. Due to the vast amounts of system and performance data needed to accurately capture their status, supercomputers require complex methods to monitor, maintain, and optimize. Data visualization is a powerful technique for overseeing these large streams of data in an easily interpretable way. The MIT Lincoln Laboratory Supercomputing Center (LLSC) enables effective monitoring through combining 3D gaming technology with compound data streams in the TX-Digital Twin, a 3D simulation of the supercomputer. The TX-Digital Twin offers both live and historical data, in visual and text formats, and tracks a multitude of revealing performance metrics.
Recent increasing interest in GPU-accelerated computing has driven a need for monitoring and maintenance of GPU-accelerated resources in supercomputers. In this paper, we build on our previous solution by integrating the visualization of additional GPU metrics, such as GPU memory usage, temperature, and power draw, into the TX-Digital Twin. Using techniques in draw call optimization, we add clear and effective displays of the new metrics while keeping the effects on performance minimal.
\end{abstract}


\section{Introduction}
With the recent rise of AI and big data, as well as persistent research in data-heavy simulation fields, GPU-accelerated computing has become a crucial asset to High Performance Computing (HPC) centers. A growing number of HPC systems are integrating GPU accelerators: in 2025, 237 of the HPC systems ranked on Top500 use accelerators or co-processors, of which most are GPUs, up from 194 a year prior \cite{top500}. This trend is not surprising—GPU acceleration offers speedups, sometimes by an order of magnitude, in a variety of applications. In 2009, a paper at the ICML conference \cite{raina2009gpu} reported speedups of up to 70x when training a deep belief network on a GPU compared to training on a dual-core CPU. Running polymer physics applications on a GPU was found to be 12.5 times faster than running them on a CPU \cite{glaser}. Supercomputing systems integrating GPU acceleration found significant gains in both throughput and power efficiency \cite{llnl_sierra}.

The MIT Lincoln Laboratory Supercomputing Center (LLSC) is an interactive, on-demand supercomputing system, with systems of over 40,000 processor cores capable of computing $10^{15}$ operations per second  \cite{llsc_system, llsc_cores}. The LLSC offers users a diverse portfolio of compute node architectures and platforms that support various workloads, including both multi-core nodes for processing-intensive computational workloads and nodes equipped with GPU accelerators for AI/ML workloads \cite{supercloudHardware}. The rapid rise and development of Large Language Models (LLM) and research surrounding training, augmentation, and performance of LLMs has demanded more powerful accelerators; as a result, the LLSC recently integrated 318 HPE GPU-accelerated nodes composed of dual AMD 9254 24-core processors with dual NVIDIA H100 NVL GPUs. The new system, TX-GAIN, premiered on the June 2025 Top500 list at 114, making it one of the largest systems in the New England\cite{LLSCTop500}. These nodes are now one of the LLSC’s most valuable resources, so monitoring and maintaining them is crucial to running the center efficiently.

\section{Existing Performance Data Pipeline}
\subsection{Data Ingestion and Scaling }
\label{Data Collection} The volume of data collected during the operation of an HPC system can overwhelm standard data aggregation methods. To this end, the LLSC developed a Dynamic Distributed Dimensional Data Model (D4M) Database~\cite{HPEC-D4M,HPEC-database} utilizing Accumulo~\cite{accumulo,HPEC-D4M-accumulo} for the back-end insertion. This system has proven to be scalable, performing efficiently even when ingesting over seven billion structured data sets.

Apache Accumulo, combined with D4M, provides a stable, performance-optimized sparse database to store the persistent raw data from the system collectors. The LLSC is then able to analyze the billions of records to observe patterns, allowing for future early detection of potential hazards.

\subsection{Data Conditioning }
The LLSC's approach to data conditioning is designed to accommodate certain functionalities important to handling large streams of data and to future changes in the computing center. Firstly, the solution should produce a structured data set that can be queried, alerted against, and correlated to user jobs. Secondly, it should support the ability to change and expand the data gathered, adjust alert thresholds, and alter methods of observation. Lastly, the data stream should be reduced to only data relevant to poor user or system performance and job failures \cite{3ddigitaltwin}. 

Due to these design criteria, the LLSC's approach to data conditioning is as follows. First, collectors (Telegraf, Slurm) and custom shell scripts gather node and user data. Additionally, Modbus registers gather environmental data like humidity, airflow, and temperature from the facility (EcoPod). Raw data are then piped through an interim Database (Influx), and custom queries pull relevant data into adaptable delimited text files. The data are ingested into Accumulo through MATLAB/D4M scripts, formatted for display, and analyzed for alerts and component changes. Lastly, data in the form of CSV or TSV files are generated for Unity\textsuperscript{TM} Gaming Engine to read and incorporate into a visual display.
\subsection{Data Display }
\begin{figure}[H]
    \centering
    \includegraphics[width=0.9\linewidth]{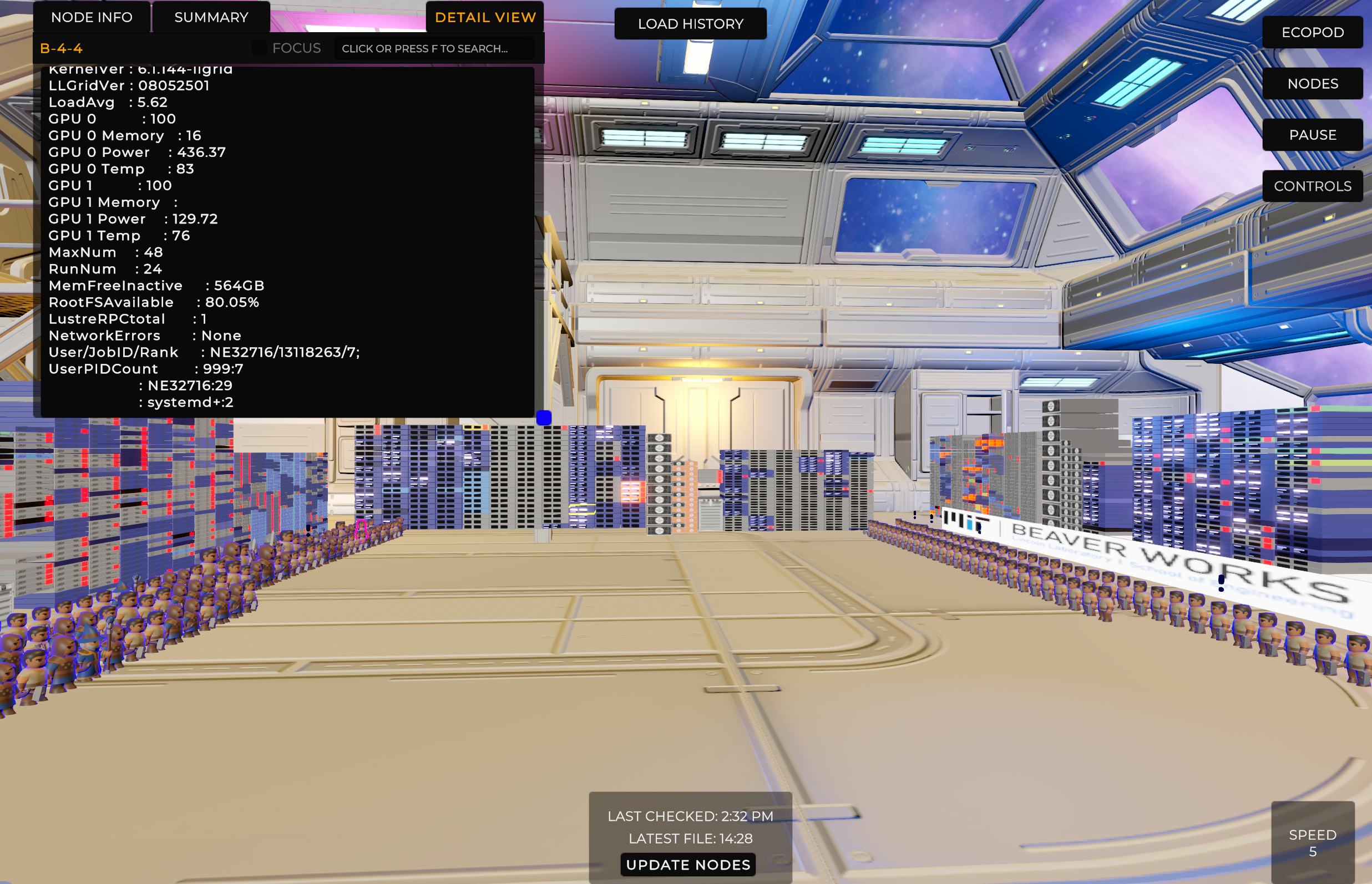}
    \caption{TX-Digital Twin full system view with 3D Graphics and text}
    \label{fig:NewVersion}
\end{figure}

To allow system administrators to efficiently parse and act on the large quantity of data, the LLSC has leveraged 3D gaming technology to visually display near-real-time data. Built in Unity\textsuperscript{TM}, the TX-Digital Twin, shown in Fig. \ref{fig:NewVersion}, collects system data, warns of alerts based on pre-set triggers, tracks users and jobs, and reports memory system usage.
The visual display of these data allows system administrators to easily identify alerts, inefficient system use, high loads, and user activity to diagnose issues with hardware, scheduling, or jobs. The TX-Digital Twin further allows administrators to track down concerns with a search tool that queries collected information and a focus mode that isolates all nodes relevant to a query or user. Additionally, the TX-Digital Twin retains historical data, so administrators can reconstruct system failures and analyze trends.

The TX-Digital Twin is a powerful tool for HPC monitoring. However, since the rise in GPU accelerators is fairly new, the tool had minimal support for tracking GPU status in-game, only tracking the GPU utilization. More sophisticated tracking of GPU-accelerated nodes would require collecting and integrating more GPU metrics.

\section{New Visual Design}
\begin{figure}[H]
    \centering
    \begin{minipage}{0.49\textwidth}
        \centering
        \includegraphics[width=0.32\linewidth]{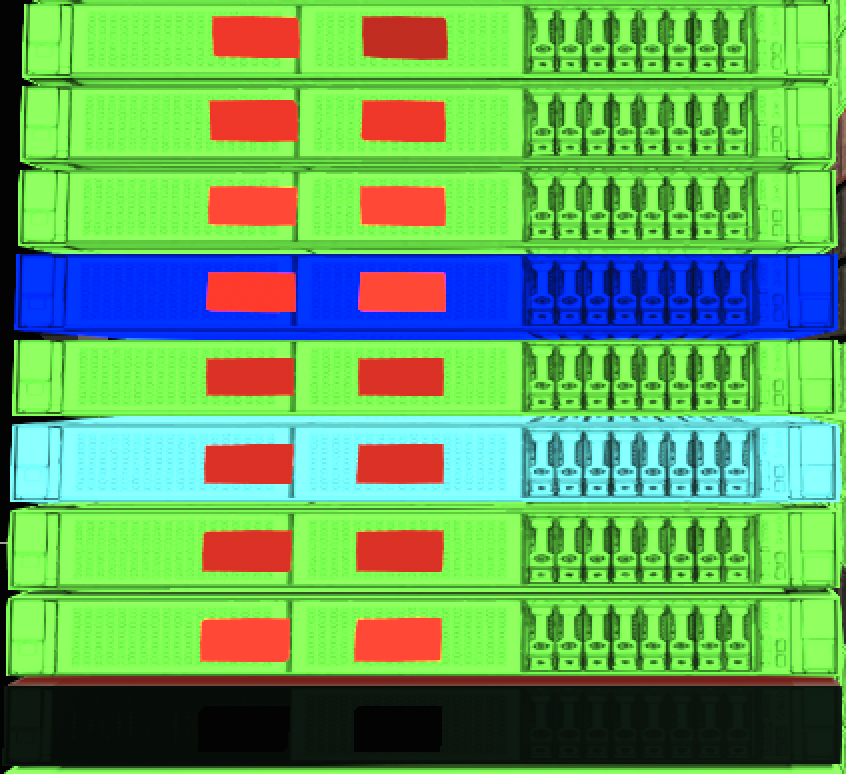}
        \includegraphics[width=0.32\linewidth]{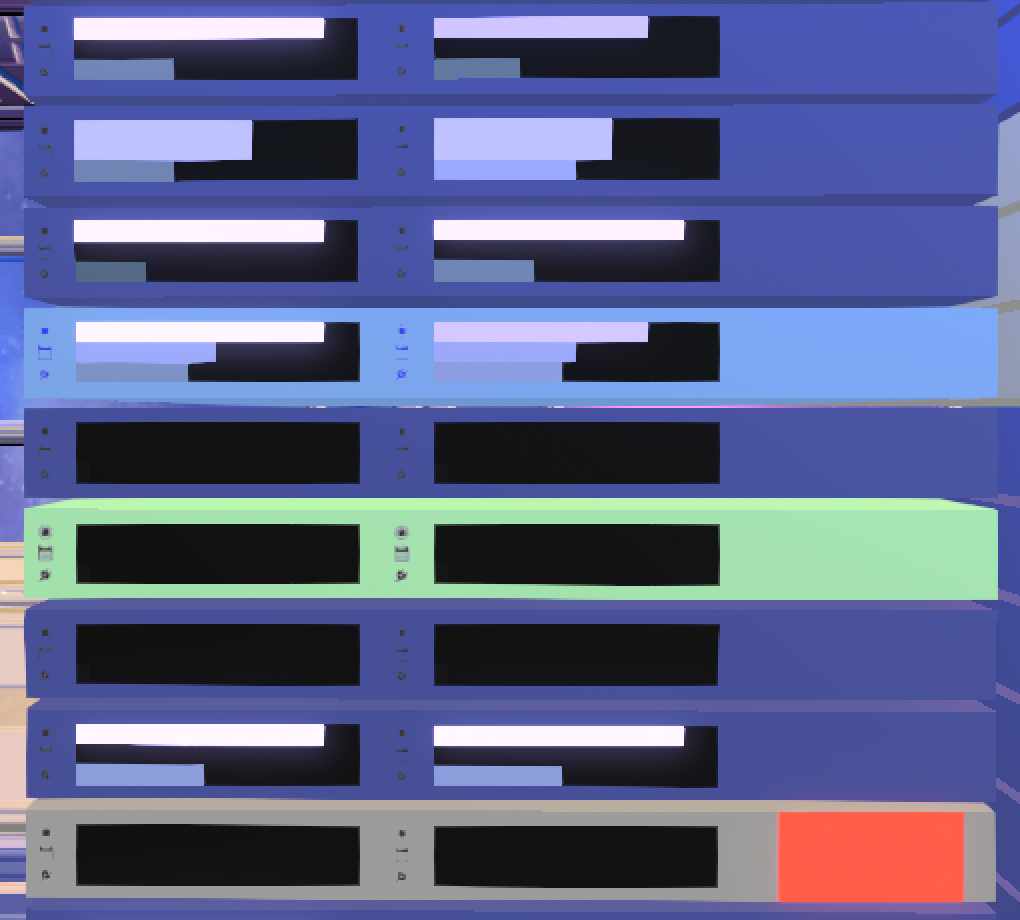}
        \includegraphics[width=0.32\linewidth, height=0.28\linewidth]{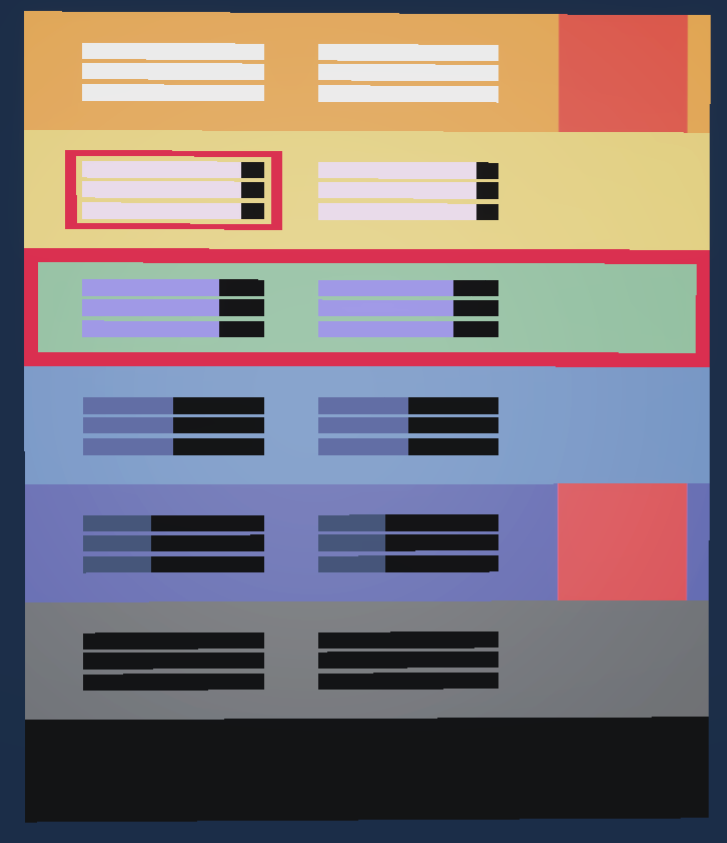}
    \end{minipage}
    \caption{In-game stack of GPU-accelerated nodes in v0.6.5 (left) compared to v0.7.2 (middle); simulated stack of nodes showing varying metric levels (right).}
    \label{fig:GPU Display Comparison}
\end{figure}

The first step in enhancing the GPU visualization was determining the most informative GPU metrics to display. In addition to GPU utilization, the TX-Digital Twin now displays GPU memory usage, GPU power draw, and GPU temperature, as shown in Fig. \ref{fig:GPU Display Comparison}. Furthermore, the new display incorporates the entire node's temperature. The existing data pipeline is well-adapted for handling expansions of the dataset, and the change in latency due to incorporating new GPU metrics, as described in Section II, was negligible. The more pressing challenge was displaying the information within the 3D Digital Twin, both from a design and performance standpoint.

A display having too many competing signals makes it hard for users to process them. With nearly 650 new GPUs, each displaying a variety of metrics, crafting an easily legible design to showcase such volumetric data is a challenge.

There are many considerations to keep in mind when designing for this high-signal environment. Firstly, metrics should be easy to read from far away. So, they should be simple enough that only a few pixels are necessary to communicate the bulk of the information. Additionally, they should be eye-catching only when they have notable information: when the metric is unusually elevated, or discordant with metrics it should track with. When the value is not notable, the metric should fade into the environment. 

We examined over a dozen possible designs, simulating them on randomized data with hundreds of nodes to evaluate their clarity. The most promising design used long, rectangular, horizontal progress bars to display GPU utilization, memory usage, and power draw, as depicted in Fig. \ref{fig:GPU Display Comparison}. 

This rectangular shape maintains integrity at a distance. The long, horizontal orientation allows for marginal changes in metric value to occupy more visual space, which increases the contrast between differing values from far away. The metric value is communicated both by the color of the bar and horizontal progress—the redundancy of information makes it easier to parse when surrounded by other signals.

For the GPU and node temperatures, we used a red outline that is only visible when the temperature surpasses tolerance, as knowing the temperature otherwise is not crucial. This reduced visual clutter while maintaining the effectiveness of the signal.

In adding the new metrics, special attention was given to colors. We restricted red to the most high-impact signals, because it most clearly signals high alert status. Here, we use red for alerts, hot components, and excessive power draw. For the node bases, we set the colors to form an intuitive gradient, with dark blue representing low CPU load and glowing orange representing high CPU load. Lastly, for the GPU metrics, using a full-color gradient would contrast poorly with some node colors, causing the signals to conflate. We instead used a gradient starting at dark purple and ending at white, so high measurements would stand out against any node color.

\section{Implementation}
Adding three dynamic progress bars per GPU into the 3D environment significantly increased the amount of geometry in the scene, which tends to increase render times and decrease performance \cite{mingeometryunity}. Since the TX-Digital Twin is used by administrators to diagnose problems in real-time, a high frame rate is important for effectiveness as a monitoring tool.

When optimizing render performance in high data environments, one primary metric is the number of draw calls, which are operations that facilitate the render process and often are expensive \cite{unity_optimizing_guide}. For each draw call, the CPU must complete a sequence of computations. A lot of these operations are fixed per call and happen regardless of the draw call size \cite{realtimerendering}. As a result, minimizing the number of draw calls or maximizing geometry per call can improve performance.

The part of the draw call that the GPU handles is computing pixel values for render targets based on the bindings and buffers that are passed in. If there is a change in material or shader, the GPU may flush and reconfigure its pipeline, so changing shader programs or material parameters between calls can be expensive \cite{realtimerendering}. Therefore, minimizing switches in materials or shaders is helpful for performance.

Recently, Unity\textsuperscript{TM} developed integrated support for increasingly powerful rendering techniques, which we took advantage of when optimizing the new version of TX-Digital Twin. Among these are a variety of batching methods, which reduce the number of draw calls or reduce the cost of draw calls, if objects in the scene match certain requirements.

There are four common batching methods supported by Unity\textsuperscript{TM}: static batching, dynamic batching, SRP batching, and GPU instancing. The speedups of each technique are highly dependent on the situation; however, for scenes with many identical objects, GPU instancing is often the most promising approach. Oftentimes, the GPU cost associated with objects using GPU instancing is higher than with other methods, but for situations with many objects, the cost can be outweighed \cite{instancingbook, instancingpaper, realtimerendering}. The TX-Digital Twin has thousands of instances of the same mesh, making it a good candidate for GPU instancing.

GPU instancing has strict requirements for implementation: using the same mesh and material across batched objects. Notably, sharing the same material means objects must have the same shader variant, keywords, per-material properties, and textures, but they can differ in hybrid per-instance shader properties, like per-instance color \cite{unitydocinstancing}. This has two implementation consequences. First, to allow a material to visually respond to data changes, the materials should be made through a common, custom shader. Second, we needed to alter the node render process to allow for instancing with per-instance property control.

\subsection{Developing Shaders}
To create an architecture capable of supporting responsive materials, we developed four instancing-compatible custom shaders using Unity\textsuperscript{TM}’s Shader Graph to handle rendering node bases, memory and load progress bars, and power draw progress bars, as well as support a toggleable outline for indicating high temperature.

The node base shader takes in a per-material input \texttt{BaseMap}, which is mixed with a calculated color map. The per-instance inputs toggle a gray base color if the node is idle; a black base color if it is off; and activate a red strip if the node has alerts. The per-instance float \texttt{Load}is used to sample the node base color from a gradient.

The GPU memory and load progress bar shader takes in a float \texttt{Load}. Based on this input, it samples a color from a gradient and masks the progress bar such that the size of the colored area corresponds to the memory use or load value.

The GPU power draw bar shader has per-material float properties of \texttt{Max} and \texttt{Min} to calibrate the progress bar to the expected power draw of the hardware. The per-instance display of the progress bar is determined by a float \texttt{Load}, which samples a color from a gradient and masks the remainder of the bar black. Any part of the \texttt{Load} exceeding the per-material float \texttt{NormalizedLarge} is colored red.

Lastly, the outline shader takes in per-material parameters that determine the proportions of the outline. The outline display is activated through the per-instance float \texttt{OutlineEnabled}.

\subsection{Code Implementation}
The structure of the TX-Digital Twin code imposes many conditions on approaches to efficient rendering. The most important requirement is per-instance control over the components of a node, to display the values of metrics. In the previous version, this was done by assigning a unique material to each node, which meant nodes typically could not be batched together. The new program, then, requires a structure that allows for per-instance overrides while maintaining a common material. Additionally, the solution needs to be compatible with ray-casting, since users should have the ability to click on individual nodes to trigger the display of corresponding detailed data. Further, nodes with different casings should have unique base textures. Lastly, the rendering process should work intuitively with the existing code structure.

In light of these considerations, between a per-instance data approach (\texttt{MaterialPropertyBlock}) and a manual instancing approach (\texttt{DrawMeshInstanced}), we went with the former. Although the latter approach would likely be more performant, it would require handling ray-casting and node rendering separately. Additionally, the manual batching and material-based grouping would require convoluted data tracking and cause a disconnect with the object-based approach used in the rest of the code.

The new code flow for updating the visuals is as follows: on initialization of the scene, we make a dictionary \texttt{AllNodes} which maps node names to node data. We read and write metric data from this dictionary. We also populate three \texttt{RenderList}s, each a dictionary for a specific node type mapping node names to structs containing the renderers and transforms of their subcomponents. In the previous version of the code, GameObjects and their renderers were often located using \texttt{Find()}; caching the frequently-accessed renderers avoids those expensive calls and reduces garbage collection.

For live updates, a delimited file containing new data is read into Unity\textsuperscript{TM}, and the changes in the data are reflected in \texttt{AllNodes}. Then, we iterate over all the \texttt{RenderList}s. For each struct containing node subcomponent renderers, we create the corresponding \texttt{RenderPropertyUpdate} struct, which specifies the renderer and material index we would like to apply changes to, and a dictionary mapping property IDs to their new values. We add the \texttt{RenderPropertyUpdate} to a \texttt{ConcurrentBag}. The purpose of this structure is to allow all computations to happen safely on worker threads, since our code leverages multi-threading to handle the volume of computations associated with the large stream of metric data. Lastly, we use Unity\textsuperscript{TM} Gaming Engine's \texttt{LateUpdate} to set the \texttt{MaterialPropertyBlocks} on the main thread after all worker threads are done. We iterate over the \texttt{ConcurrentBag}; for each \texttt{RenderPropertyUpdate}, we apply the changes to a \texttt{MaterialPropertyBlock}, and push the changes with \texttt{SetPropertyBlock}.\\[-3.0ex]

\subsection{Performance Comparison}
Due to the addition of six bars per GPU-accelerated node and other mesh changes, the amount of geometry in the new version increased compared to the previous version. The change in geometry can be quantized by the number of triangles rendered per frame, which is shown in Table \ref{tab:TriBatch}, with the new version handling around 8.28 times more triangles per frame. Notably, however, the number of batches per frame has decreased by a factor of two, due to the newly implemented GPU instancing, which handles 8.8k potential draw calls per frame in only 553 batches. 

\begin{figure}
    \centering
    \includegraphics[width=0.9\linewidth]{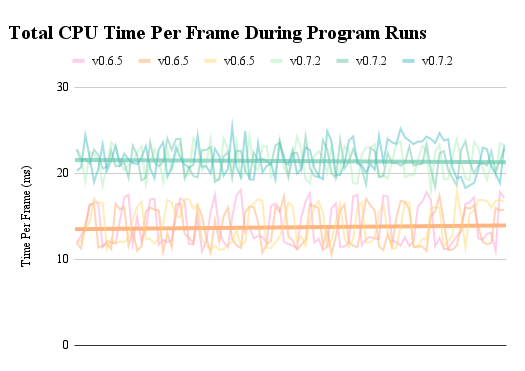}
    \caption{Total CPU Time Per Frame During Program Runs. Data represents samples from 100 frames from three runs of each version.}
    \label{fig:CPUTotal}
\end{figure}

\begin{figure}
    \centering
    \includegraphics[width=0.55\linewidth]{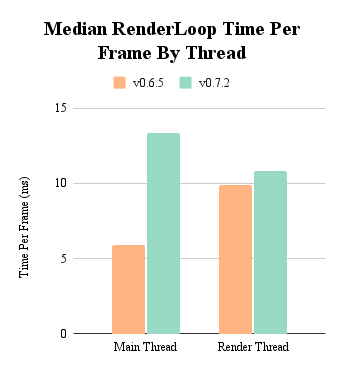}
    \caption{Median value across five independent measurements of median timing of RenderLoop on the main and render threads. }
    \label{fig:Threads}
\end{figure}

\begin{table}
    \caption{Batch and Triangle Counts per Frame by Version}
    \centering
    \begin{tabular}{|c|c|c|c|}\hline
         & \textbf{v0.6.5} & \textbf{v0.7.2} & \textbf{Multiplier} \\\hline
      \textbf{Batches}   & 3588 & 1677  & x0.47\\\hline
      \textbf{Triangles}   & 204k & 1688.5k & x8.28\\ \hline  
    \end{tabular}
    \label{tab:TriBatch}
\end{table}
Despite the decrease in the number of batches per frame, the increase in geometry did negatively affect performance, raising the average CPU time per frame from around 14ms to around 21.5ms, as shown in Fig. \ref{fig:CPUTotal}.
Narrowing in on the effects on render times, data were gathered on the length of \texttt{RenderLoop} processes on the main thread and render thread to observe the effect on CPU-side render processes and communication of render data to the GPU, respectively. Fig. \ref{fig:Threads} compares the impact of the new additions on the \texttt{RenderLoop} lengths on the main and render threads. The timing of the render thread \texttt{RenderLoop} barely increased, which suggests the new geometry has been efficiently handled by instancing and \texttt{MaterialPropertyBlock}. The length of the main thread \texttt{RenderLoop}, however, increased considerably compared to the previous version, suggesting per-frame computations of draw calls are costly. Future optimizations should target the CPU side of rendering.  

\section{Conclusion and Future Work}
The latest iteration of the LLSC TX-Digital Twin reflected in this paper achieved the goal of significantly elevating GPU visualization, monitoring, and  management capability for the LLSC. This required a considerable increase in data ingestion, processing, and in-game geometry, which was largely mitigated by coding more efficient batching and rendering while maintaining overall functionality. The expansion in situational awareness offered by the new version justifies the slight performance decrease, and we have identified methods of further improving performance in the future. 

The performance data indicate that calculations for \texttt{MaterialPropertyBlock} contribute significantly to an increased CPU overhead, and next steps towards improving performance should address the CPU bottleneck. The current implementation, with setting one property block for each node, is likely expensive; moving towards crafting manual batches and buffers using arrays of property blocks is one possibility. Using \texttt{DrawMeshInstanced}, we can set a higher number of instances per call manually, and this approach is more compatible with the SRP rendering pipeline than \texttt{MaterialPropertyBlock} is \cite{unity_mpb}. However, it would require separating the node rendering from the node \texttt{GameObjects}, as well as rewriting the display-updating portion of the program to become material-centered as opposed to object-centered. A potentially promising approach is moving towards an Entity Component System (ECS)-based structure with Hybrid Renderer. The ECS pipeline is made to handle big data and render it efficiently with instancing. However, it likewise has reduced compatibility with \texttt{GameObjects}, and would cause a major restructuring to the code. It is possible, though, that migrating over from the MonoBehavior structure the TX-Digital Twin currently uses to an ECS-based structure could unlock significant performance gains, especially as the amount of specialized data collected grows with the evolving LLSC supercomputer.

\bibliographystyle{IEEEtran}   % IEEE style
\bibliography{references}      % points to references.bib
\fontsize{7}{8}\selectfont
%%%%
\section*{Acknowledgment}
The authors wish to acknowledge the following individuals for their contributions and support: Daniel Mojica, Daniel Andersen, LaToya Anderson, Bill Arcand, Sean Atkins, Chris Berardi, Bob Bond, Alex Bonn, Bill Cashman, K Claffy, Tim Davis, Chris Demchak, Alan Edelman, Peter Fisher, Jeff Gottschalk, Thomas Hardjono, Chris Hill, Michael Houle, Michael Jones, Charles Leiserson, Piotr Luszczek, Kirsten Malvey, Peter Michaleas, Lauren Milechin, Chasen Milner, Sanjeev Mohindra, Guillermo Morales, Julie Mullen, Heidi Perry, Sandeep Pisharody, Christian Prothmann, Andrew
Prout, Steve Rejto, Albert Reuther, Antonio Rosa, Scott Ruppel, Daniela Rus, Mark Sherman, Scott Weed, Charles Yee,
Marc Zissman.
\end{document}